\documentclass{article}

\usepackage{PRIMEarxiv}

\usepackage[utf8]{inputenc} % allow utf-8 input
\usepackage[T1]{fontenc}    % use 8-bit T1 fonts
\usepackage{hyperref}       % hyperlinks
\usepackage{url}            % simple URL typesetting
\usepackage{booktabs}       % professional-quality tables
\usepackage{amsfonts}       % blackboard math symbols
\usepackage{nicefrac}       % compact symbols for 1/2, etc.
\usepackage{microtype}      % microtypography
\usepackage{lipsum}
\usepackage{fancyhdr}       % header
\usepackage{graphicx}       % graphics
\usepackage{amsmath}
\graphicspath{{media/}}     % organize your images and other figures under media/ folder

\title{Modeling the Repetition-based Recovering of Acoustic and Visual Sources with Dendritic Neurons
}

\author{
  Giorgia Dellaferrera \\
  Okinawa Institute of Science and Technology\\ 
  Okinawa, Japan; \\
  Institute of Neuroinformatics \\
  University of Zurich and ETH Zurich\\ Zurich, Switzerland; \\
  Department of Ophthalmology, Children’s Hospital\\ Harvard Medical School\\ Boston, MA, United States\\
  \texttt{giorgia.dellaferrera@childrens.harvard.edu} \\
   \And
  Toshitake Asabuki, Tomoki Fukai \\
  Okinawa Institute of Science and Technology\\ 
  Okinawa, Japan \\
  \texttt{\{toshitake.asabuki2, tomoki.fukai\}@oist.jp} \\
}

\begin{document}
\maketitle

\begin{abstract}
In natural auditory environments, acoustic signals originate from the temporal superimposition of different sound sources. The problem of inferring individual sources from ambiguous mixtures of sounds is known as blind source decomposition. Experiments on humans have demonstrated that the auditory system can identify sound sources as repeating patterns embedded in the acoustic input. Source repetition produces temporal regularities that can be detected and used for segregation. Specifically, listeners can identify sounds occurring more than once across different mixtures, but not sounds heard only in a single mixture. However, whether such a behaviour can be computationally modelled has not yet been explored.
Here, we propose a biologically inspired computational model to perform blind source separation on sequences of mixtures of acoustic stimuli. Our method relies on a somatodendritic neuron model trained with a Hebbian-like learning rule which can detect spatio-temporal patterns recurring in synaptic inputs. We show that the segregation capabilities of our model are reminiscent of the features of human performance in a variety of experimental settings involving synthesized sounds with naturalistic properties. Furthermore, we extend the study to investigate the properties of segregation on task settings not yet explored with human subjects, namely natural sounds and images.
Overall, our work suggests that somatodendritic neuron models offer a promising neuro-inspired learning strategy to account for the characteristics of the brain segregation capabilities as well as to make predictions on yet untested experimental settings.
\end{abstract}

% keywords can be removed
\keywords{dendritic neurons \and spiking neural networks \and blind source separation \and sound source repetition \and spatio-temporal structure }

\section{Introduction}
Hearing a sound of specific interest in a noisy environment is a fundamental ability of the brain that is necessary for auditory scene analysis. To achieve this, the brain has to unambiguously separate the target auditory signal from other distractor signals. In this vein, a famous example is the ``cocktail party effect'', \textit{i.e.}, the ability to distinguish a particular speaker’s voice against a
multi-talker background \cite{Brown01,Mesgarani12}. Many psychophysical and neurobiological studies have been conducted to clarify the psychophysical properties and underlying mechanisms of the segregation of mixed signals \cite{mcdermott09,Bee08,Asari06,McDermott11,lewald15,li17,Atilgan18,narayan08,schmidt11}, and computational theories and models have also been proposed for this computation
\cite{thakur15,haykin05,Kameoka18,karamatli18,sawada19,sagi01,bell95,amari95,Dong15,elhilali09}. However, how the brain attains its remarkable sound segregation remains elusive.
%This modeling study discusses two stage model, a feature analysis stage that maps the acoustic input into a multidimensional cortical representation and an integrative stage that recursively builds up expectations of how streams evolve over time. Our model does not attempt the second stage, which I think describes a limitation of our model. So, the arguments shown this manuscript may be useful for shaping the discussion in our manuscript. CHECK
Various properties of auditory cues such as spatial cues in binaural listening \cite{ding12} and temporal coherence of sound stimuli \cite{krishnan14,teki13} are known to facilitate the listener’s ability to segregate a particular sound from the background. Auditory signals that reached to ears first undergo the analysis of frequency spectrums by cochlea \cite{oxenham18}. Simultaneous initiation and termination of the component signals and the harmonic structure of the frequency spectrums help the brain to identify the components of the target sound \cite{popham18}. Prior knowledge about the target sound, such as its familiarity to listeners \cite{Woods18,Elhilali13}, and top-down attention can also improve their ability to detect the sound \cite{Ahveninen11,ZionGolumbic13,Bronkhorst15,Xiang10,osullivan14,kerlin10}. However, many of these cues are subsidiary and not absolutely required for hearing the target sound. For example, a mixture sound can be separated by monaural hearing \cite{Hawley04} or without spatial cues \cite{Middlebrooks20}. Therefore, the crucial mechanisms of sound segregation remain to be explored.  

Whether or not biological auditory systems segregate a sound based on principles similar to those invented for artificial systems remains unclear \cite{mcdermott09,Bee08}. Among such principles, independent component analysis (ICA) \cite{Comon94} and its variants are the conventional mathematical tools used for solving the sound segregation problem, or more generally, the blind source decomposition problem \cite{haykin05,bell95,amari95,hyvarinen97}. Owing to its linear algebraic features, the conventional ICA requires as many input channels (e.g., microphones) as the number of signal sources, which does not appear to be a requirement for sound segregation in biological systems. It has been suggested as an alternative possibility that human listeners detect latent recurring patterns in the spectro-temporal structure of sound mixtures for separating individual sound sources \cite{McDermott11}. This was indicated by the finding that listeners could identify a target sound when the sound was repeated in different mixtures in combination with various other sounds but could not do so when the sound was presented in a single mixture. 

The finding represents an important piece of information about the computational principles of sound source separation in biological systems. Here, we demonstrate that a computational model implementing a pattern-detection mechanism accounts for the characteristic features of human performance observed in various task settings. To this end, we constructed a simplified model of biological auditory systems by using a two-compartment neuron model recently proposed for learning regularly or irregularly repeated patterns in input spike trains \cite{asabuki20}. Importantly, this learning occurs in an unsupervised fashion based on the minimization principle of regularized information loss. The unnecessity of teaching signals has an important implication that the essential computation of sound source segregation can emerge at the single-neuron level. Furthermore, it was previously suggested that a similar repetition-based learning mechanism may also work for the segregation of visual objects \cite{McDermott11}. To provide a firm computational ground, we extended the tasks of our framework to predictions on visual images.

%\section{Article types}

%For requirements for a specific article type please refer to the Article Types on any Frontiers journal page. Please also refer to  \href{http://home.frontiersin.org/about/author-guidelines#Sections}{Author Guidelines} for further information on how to organize your manuscript in the required sections or their equivalents for your field

% For Original Research articles, please note that the Material and Methods section can be placed in any of the following ways: before Results, before Discussion or after Discussion.

\section{Results}
\subsection{Learning of repeated input patterns by a two-compartment neuron model }
We used a two-compartment spiking neuron model which learns recurring temporal features in synaptic input. In short, the dendritic compartment attempts to predict the responses of the soma to given synaptic input by modelling the somatic responses. To this end, the neuron model minimizes information loss within a recent period when the somatic activity is replaced with its model generated by the dendrite (see Material and Methods for details). The prediction is learnable when input spike sequences from presynaptic neurons are non-random and contain recurring temporal patterns. In such a case, the minimization of information loss induces a consistency check between the dendrite and soma, eventually enforcing both compartments to respond selectively to one of the patterns. Mathematically, the somatic response serves as a teaching signal to supervise synaptic learning in the dendrite. Biologically, backpropagating action potentials may provide the supervising signal \cite{Larkum13,Larkum99}. 

\begin{figure}[!h]
 \centering
     \includegraphics[width=1.\textwidth]{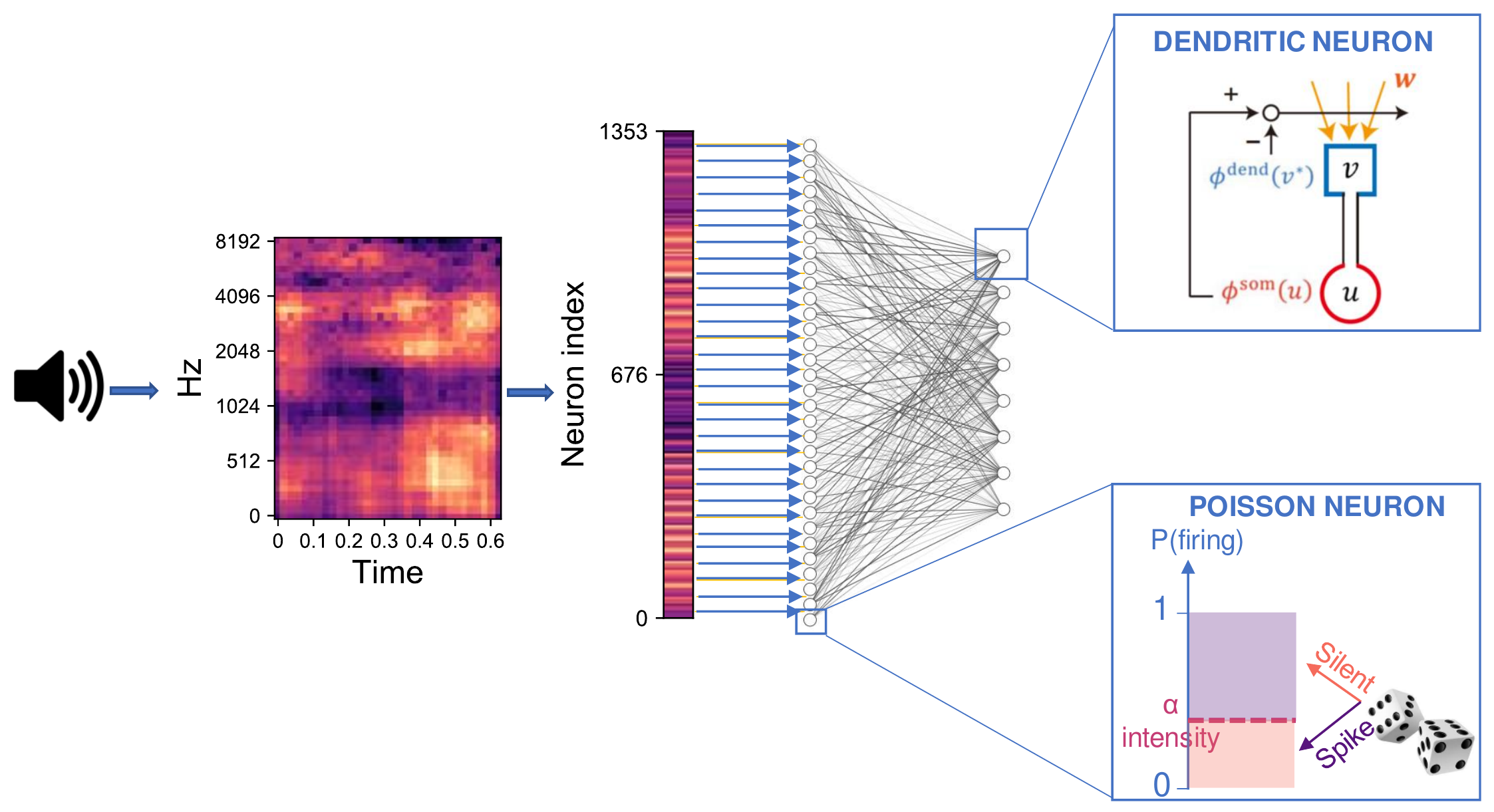}
\caption{{\bf Network architecture.}
The input signal is pre-processed into a two-dimensional image (\textit{i.e.}, the spectrogram) with values normalized in the range [0,1]. The image is flattened into a one-dimensional array where the intensity of each element is proportional to the Poisson firing activity of the associated input neuron. The neurons in the input layer are connected to those in the output layer through either full connectivity or random connectivity with connection probability p = 0.3. The output neurons are trained following the artificial dendritic neuron learning scheme \cite{asabuki20}.}
\label{fig1}
\end{figure}

We constructed an artificial neural network based on the somatodendritic consistency check model and trained the network to perform the task of source recovering from embedded repetition. The network consisted of two layers of neurons. The input layer encoded the spectrogram of acoustic stimuli into spike trains of Poisson neurons. The output layer was a competitive network of the two-compartment models that received synaptic input from the input layer and learned recurring patterns in the input (figure \ref{fig1}). We designed the output layer and the learning process similarly to the network used previously \cite{asabuki20} for the blind signal separation (BSS) within mixtures of multiple mutually correlated signals. In particular, lateral inhibitory connections between the output neurons underwent spike-timing-dependent plasticity for self-organizing an array of feature-selective output neurons (Material and Methods). 

\subsection{Synthesized and natural auditory stimuli }
We examined whether the results of our computational model are consistent with the outcomes of the experiments on human listeners on artificially synthesized sounds described previously \cite{McDermott11}. To provide a meaningful comparison with the human responses, we adopted for our simulations settings as close as possible to the experiments, both in terms of dataset generation and performance evaluation (Material and Methods). In the human experiments, a dataset containing novel sounds was generated such that listeners’ performance in sound source segregation was not influenced by familiarity with previously experienced sounds. To closely reproduce the experiment, we created a database of synthesized sounds according to the same method as described in \cite{McDermott11} (Material and Methods). The synthesized stimuli retained similarity to real-world sounds except that they lacked grouping cues related to temporal onset and harmonic spectral structures. Furthermore, unlike human listeners, our neural network was trained and built from scratch, and had no previous knowledge of natural sounds that could bias the task execution. We exploited this advantage to investigate whether and how the sound segregation performance was affected by the presence of grouping cues in real sounds. To this goal we also built a database composed of natural sounds (Material and Methods). 

\begin{figure}[!h]
 \centering
     \includegraphics[width=1.\textwidth]{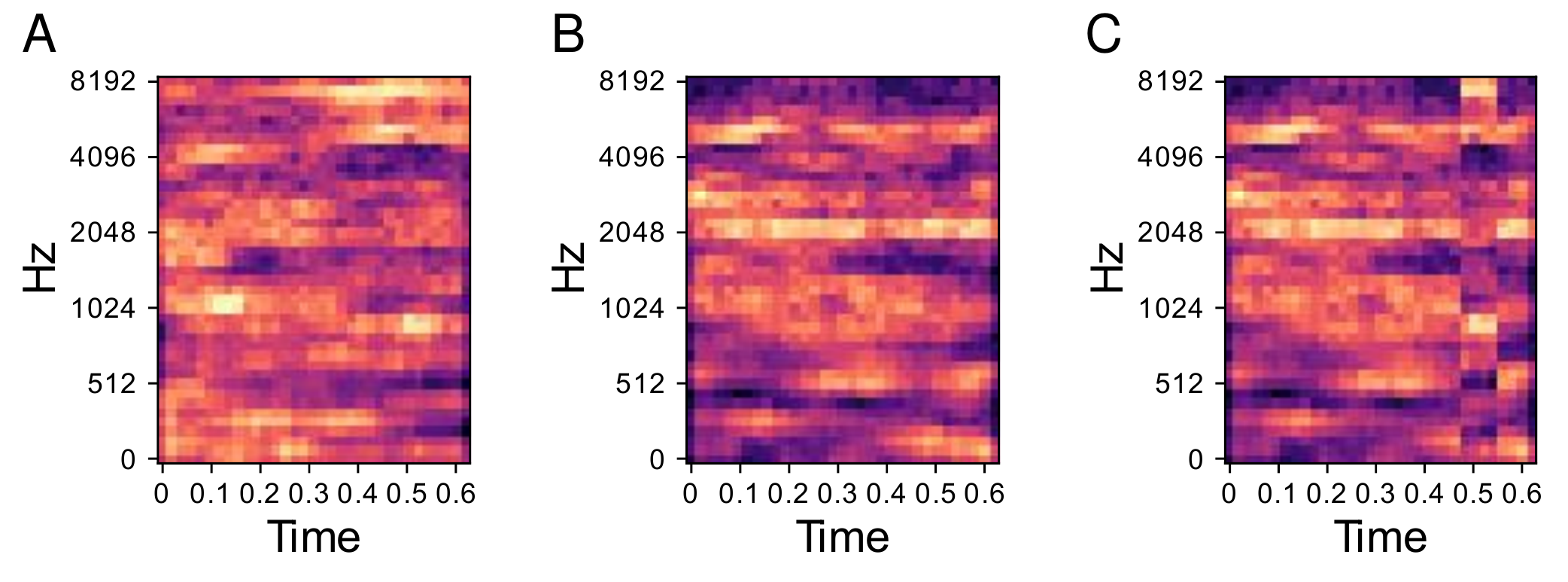}
\caption{{\bf Synthesized sounds - Target and associated distractor.}
(\textbf{A}) Spectrogram of one target sound. (\textbf{B}) Step 1 to build the spectrogram of an incorrect probe related to the target in A: a sound is randomly selected from the same Gaussian distribution generating the target. (\textbf{C}) Step 2 to build the  incorrect probe: after the sampling, a randomly selected time slice equal to 1/8 of the sound duration is set to be equal to the target. In the figure, the temporal slice is the vertical stripe around time 0.5s.}
\label{fig2}
\end{figure}

To build the sequence of input stimuli, we randomly chose a set of sounds from the database of synthesized or natural sounds, and we generated various mixtures by superimposing them. We refer to the main sound, which is always part of mixtures, as the \textit{target}, and to all the other sounds, which were either presented as mixing sounds with the target (\textit{i.e., masker sounds}) or presented alone, as \textit{distractors}. The target sound is shown in red in the training protocols (figure \ref{fig5} and figure \ref{fig6}). Following the protocol in \cite{McDermott11}, we concatenated the mixtures of target and distractors into input sequences. For certain experiments, we also included unmixed distractor sounds. We presented the network with the input sequence for a fixed number of repetitions. During the input presentation, the network's parameters evolved following the learning rule described in \cite{asabuki20}. Then, we examined the ability of the trained network to identify the target sound (as well as the masker sounds) by using probe sounds, which were either the target or distractor sound composing the mixtures presented during training (\textit{correct probe}) or a different sound (\textit{incorrect probe}). Incorrect probes for synthesized target sounds were generated by using the same method as described in \cite{McDermott11}.  Examples of target sounds, distractor sounds and incorrect probes are shown in figure \ref{fig2}A, figure \ref{fig2}B and figure \ref{fig2}C, respectively. 

\subsection{Learning of mixture sounds in the network model }
\begin{figure}[!h]
 \centering
     \includegraphics[width=1.\textwidth]{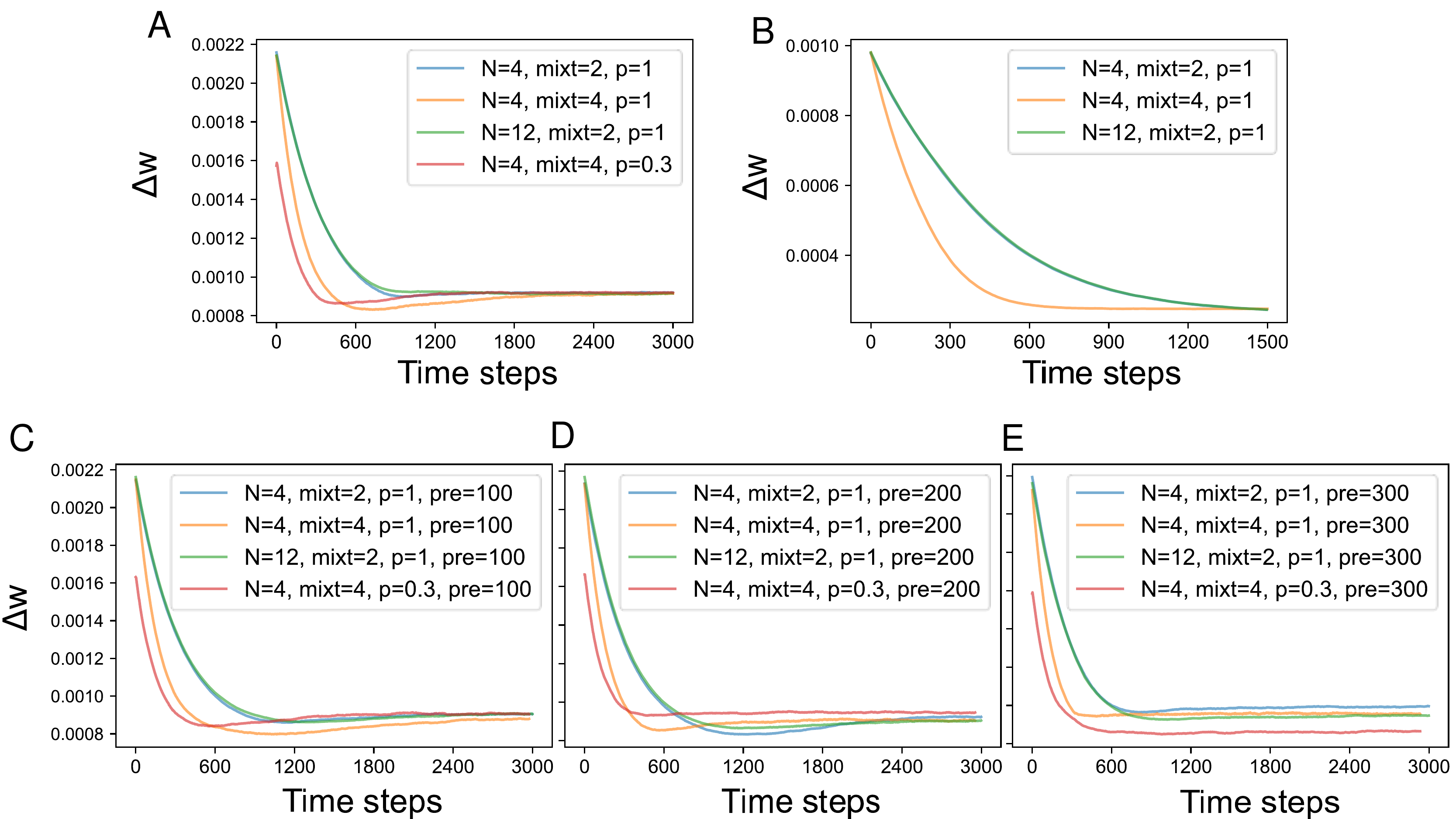}
\caption{{\bf Learning curves.}
(\textbf{A}) Average synaptic weight change for the experiments carried out on the synthetized sounds, the network being initialized with random values. (\textbf{B}) Average synaptic weight change for the experiments carried out on the natural sounds, the network being initialized with random values.  (\textbf{C}) Average synaptic weight change for the experiments carried out on the synthetized sounds, the network being pretrained on the targets set presented for 100 epochs.  (\textbf{D}) Average synaptic weight change for the experiments carried out on the synthetized sounds, the network being pretrained on the targets set presented for 200 epochs.  (\textbf{E}) Average synaptic weight change for the experiments carried out on the synthetized sounds, the network being pretrained on the targets set presented for 300 epochs.  
In all cases, when the number of output neurons is varied no significant change is found, please refer to the blue and green curves showing respectively the training with 4 and 12 output neurons. Furthermore, the figures show that both when a larger number of training mixtures is presented (yellow curves) and when only 30\% of the connections are kept (red curves) the slope of the learning curve is steeper.
}
\label{fig3}
\end{figure}

Our network model contained various hyperparameters such as number of output neurons, number of mixtures and connectivity pattern. A grid search was performed to find the optimal combination of hyperparameters. Figure \ref{fig3}A and figure \ref{fig3}B report the learning curves obtained on synthesized and natural sounds, respectively, for random initial weights and different combinations of hyperparameters. For both types of sounds, synaptic weights changed rapidly in the initial phase of learning. The changes were somewhat faster for synthesized sounds than for natural sounds, but the learning curves behaved similarly for both sound types. Number of output neurons little affected the learning curves, while they behaved differently for different connectivity patterns or different numbers of mixtures. Because familiarity to sounds enhances auditory perception in humans \cite{Jacobsen05}, we investigated whether pretraining with a sequence containing target and distractors improves learning in our model for various lengths of pretraining. Neither the training speed nor the final accuracy were significantly improved by the pretraining (figure \ref{fig3}C, figure \ref{fig3}D and figure \ref{fig3}E).

\begin{figure}[!h]
 \centering
     \includegraphics[width=1\textwidth]{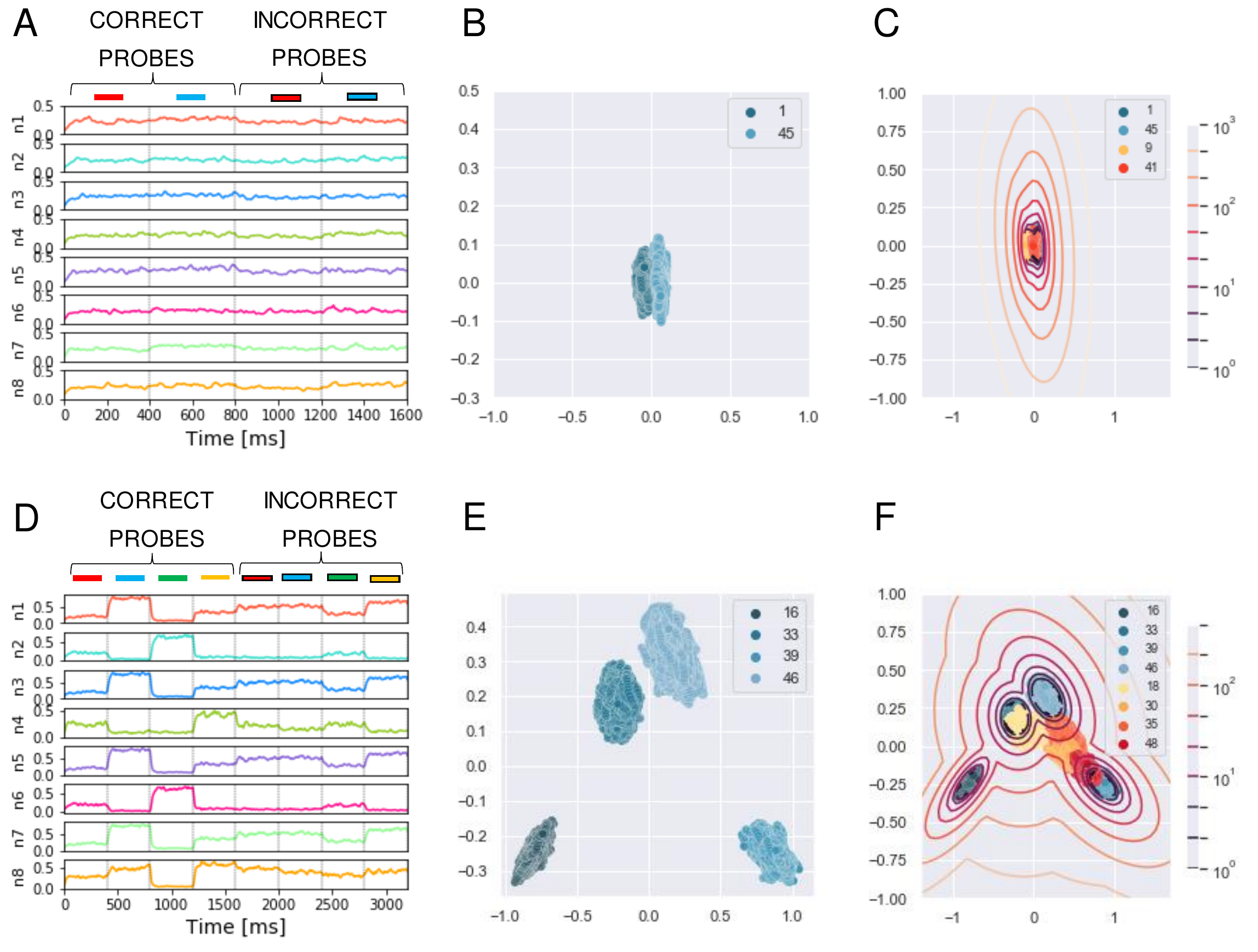}
\caption{{\bf Experiment 1 - output dynamics and clustering.}
\textbf{(A,B,C}) refer to the results of Experiment 1 on synthesized sounds with a single mixture presented during training. (\textbf{D,E,F}) refer to the results of Experiment 1 on synthesized sounds with three mixtures presented during training. (\textbf{A}) Voltage dynamics of the 8 output neurons during inference, when the target, the distractor and the two associated incorrect probes are tested. As expected, the neuron population is not able to respond with different dynamics to the four sounds, and the voltage of all the output neurons fluctuates randomly throughout the whole testing sequence. (\textbf{B}) The PCA projection of the datapoints belonging to the two targets (in blue) shows that the clusters are collapsed into a single cluster. (\textbf{C}) When GMM is applied, all the datapoints representing both the correct probes (in blue) and the incorrect probes (in orange and red) fall within the same regions, making it impossible to distinguish the different sounds based on the population dynamics. (\textbf{D}) Voltage dynamics of the 8 output neurons during inference, when the four targets and the associated distractors are tested. As expected, the neuron population has learnt the feature of the different sounds and responds with different dynamics to the eight sounds. Each output neuron exhibits an enhanced response to one or few sounds. (\textbf{E}) The PCA projection of the datapoints belonging to the four correct probes (in blue) shows that the clusters are compact and spatially distant one from the other. (\textbf{F}) When GMM is applied, the model shows that the network is, most of the times, able to distinguish the target and distractors (in blue) from the incorrect probes (in yellow, orange and red). The targets are never confused among each other. Three of the four distractors fall far from the targets’ region, while the fourth (in yellow) overlaps with one of the targets. These results are overall coherent with the human performance.}
\label{fig4}
\end{figure}

To reliably compare the performance of our model with human listeners, we designed a similar assessment strategy to that adopted in the experiment.  In \cite{McDermott11} listeners were presented with mixtures of sounds followed by a probe which could be either a correct probe (\textit{i.e.}, the target sound present in the training mixtures) or an incorrect probe (\textit{i.e.}, sounds unseen during the training). The subjects had to say whether they believed the probe was present in the training mixture by using one of the four responses “sure no” “no”, “yes”, and “sure yes”. The responses were used to build a receiver operating characteristics (ROC) as described in \cite{wickens02}, and the area under the curve (AUC) was used as performance measure, with AUC = 0.5 and 1 corresponding to chance and perfect correct, respectively. In our algorithm, we mimicked this protocol for reporting by using the likelihood as a measure of performance. To this goal, first, for each tested probe, we projected the response of the N output neurons (figure \ref{fig4}A,D) to a two-dimensional PCA projection plane. We defined the PCA space based on the response to the correct probes and later projected on it the datapoints related to the incorrect probes (figure \ref{fig4}B,E). Second, we clustered the datapoints related to the correct probes through a Gaussian Mixture Model (GMM) with as many classes as the number of correct probes (figure \ref{fig4}C,F). Third, for each datapoint we computed the likelihood that it belonged to one of the clusters. The target likelihood values are fixed to 1 and 0 for datapoints related to correct and incorrect probes respectively. We binned the likelihood range into four intervals corresponding, in an ascending order, to the four responses “sure no”, “no”, “yes”, and “sure yes” (for additional details see Material and Methods). Now, we are ready to examine the performance of the model in a series of experiments. We show examples of the different behaviour of the network trained on single (figure \ref{fig4}A-C) or four mixtures (figure \ref{fig4}D-F). As expected, the ability of the model to learn and distinguish the targets from the distractors depended crucially on the number of mixtures.

The algorithm was implemented in Python and a sample code used to simulate Experiment 1 is available at the repository \url{https://github.com/GiorgiaD/dendritic-neuron-BSS}. 

\subsection{Experiment 1: Sound Segregation with Single and Multiple Mixtures of Synthesized Sounds }
\begin{figure}[t]
 \centering
     \includegraphics[width=1\textwidth]{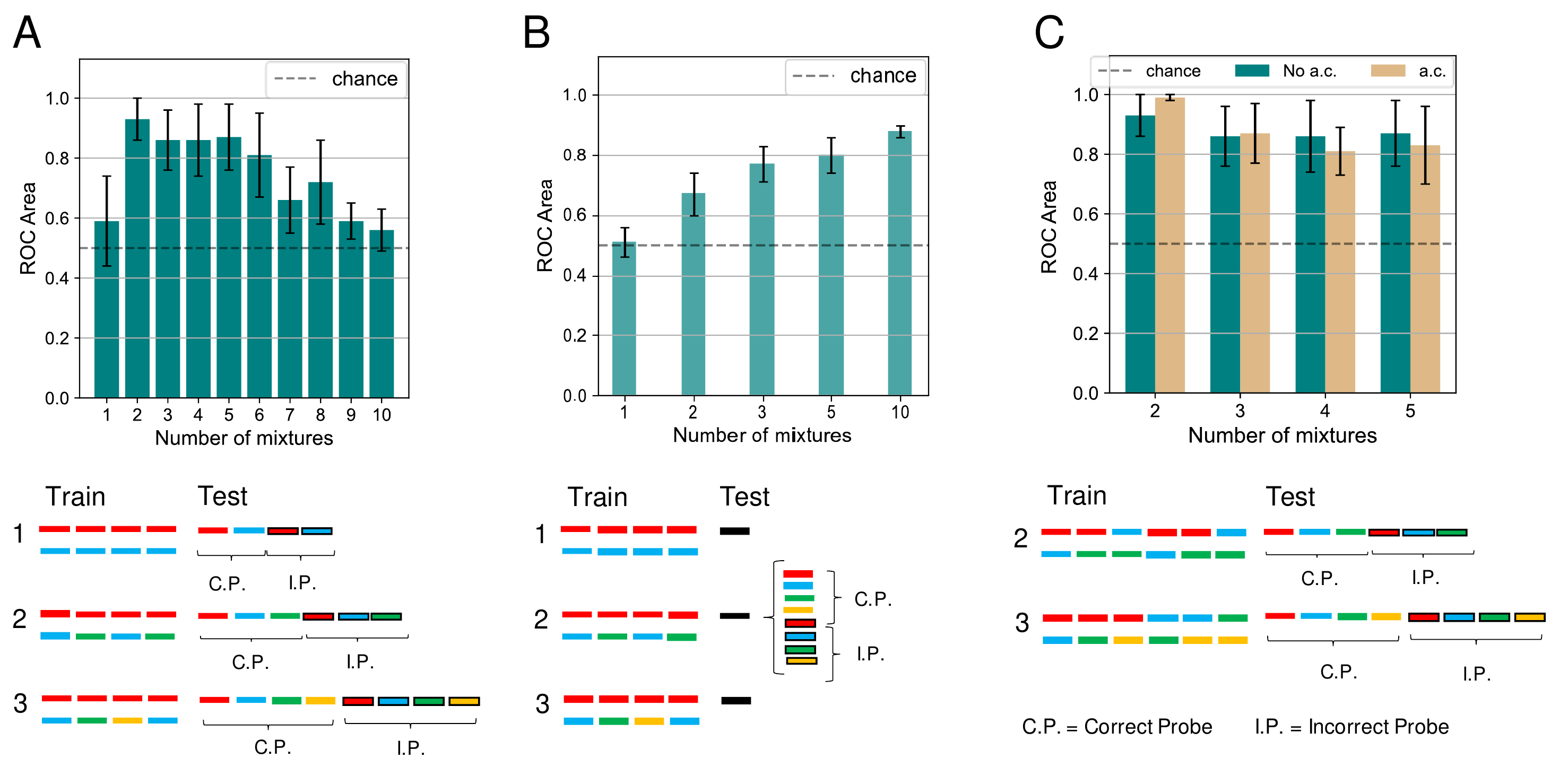}
\caption{{\bf Experiment 1 and 1 a.c. – results and comparison with human performance.}
(\textbf{A}) Results and schematics for Experiment 1 on the dendritic network model. The number of mixtures is varied from 1 to 10. Performance is close to chance for a single training mixture. The performance is boosted as two mixtures are presented. As the number of mixtures is further increased, the clustering accuracy slowly decreases towards chance values. (\textbf{B}) Results and schematics for Experiment 1 on the human experiment. The number of mixtures presented are 1,2,3,5 and 10. For a single mixture the performance is close to chance. As the number of mixtures increases, the classification accuracy improves steadily. (\textbf{C}) Results and schematics for Experiment 1 a.c. on the dendritic network model. The number of mixtures is varied from 2 to 5. Combining all the mixing sounds in mixtures slightly improves the mean performance for two mixing sounds, while it slightly worsens it for a larger number of mixtures.}
\label{fig5}
\end{figure}

To begin with, we compared how the number of mixtures influences the learning performance between human subjects and the model. The number of mixtures presented during training was varied from 1, where no learning was expected, to 2 or more, where the model was expected to distinguish the target sounds from their respective distractors. The simulation protocol is shown in figure \ref{fig5}A (bottom).  
As reported in figure \ref{fig5}A (top), we obtained that, when one mixture only was shown, neither the target nor the mixing sound was learnt, and performance was close to chance. An immediate boost in the performance was observed when the number of mixtures was raised to two. The network managed to distinguish the learnt targets from the incorrect probes with an accuracy greater than 90\%. As the number of mixtures increased up to six, the accuracy worsened slightly, remaining above 80\%. A significant drop in the performance was observed for a greater number of mixtures. From a comparison with the results shown in figure \ref{fig5}B, which were replicated for human subjects \cite{McDermott11}, it emerged that our model was able to partially reproduce human performance: the success rate was at chance levels when training consists of a single mixture only; the target sounds could be distinguished to a certain accuracy if more than a mixture was learnt. We also verified that the model performance was robust for variations of the network architecture, both in terms of the number of output neurons \textit{N} and the connection probability \textit{p} (Supplementary figure 1).

Our model and human subjects also exhibited interesting differences. When the mixture number was increased to two, performance improved greatly in our model but only modestly in human subjects. Unlike human subjects, our model showed a decreasing accuracy as the number of mixtures further increased. 
We consider that such discrepancies could arise from a capacity limitation of the network. Indeed, the network architecture is very simple and consists of two layers only, whose size is limited by the spectrogram dimensions for the input layer and by the number of output neuron for the last layer. Therefore the amount of information that the network can learn and store is limited with respect to the significantly more complex structure of the human auditory system. We also suspect that the two-dimensional PCA projection might limit the model performance when a large number of distractors is used. Indeed the PCA space becomes very crowded and although the datapoints are grouped in distinct clusters, the probability that such a cluster lie close to each other is high. To verify this hypothesis, we tested a modification of the inference protocol of the algorithm. During test, we presented the network only with the target sound and one incorrect probe, and performed BSS on the PCA space containing the two sounds. Under this configuration, the model performance is above chance level for two or more different mixtures, and the accuracy does not significantly decrease for large number of mixtures (Supplementary figure 2). 

We may use our model for predicting performance of human subjects in auditory perception tasks not yet tested experimentally. To this end, we propose an extension of the paradigm tested previously: for set-ups with the number of mixtures between two and five, we investigated whether presenting all possible combinations of the mixing sounds among themselves, rather than only the distractors with the target, affects the performance. The experiment is labelled ``Experiment 1 a.c.'', where a.c. stands for ``all combinations'', and its training scheme is reported in figure \ref{fig5}C. Because all sounds are in principle learnable in the new paradigm, we expect an enhanced ability of distinguishing the correct probes from the incorrect ones. Somewhat unexpectedly, however, our model indicated no drastic changes in the performance when the mixture sequence presented during training contained all possible combinations of the mixing sounds. Such a scheme resulted in a minor improvement in the accuracy only for the experiments with two mixing sounds. Indeed, in the ``all combinations'' protocol, during training the distractor was presented in more than one different mixture, while in the original task setting only the target was combined with different sounds. We hypothesize that the ``all combinations'' protocol makes it easier for the network to better distinguish the distractor sound. For four or five mixing sounds, instead, the performance slightly worsened. It is likely that this behaviour is related to the already mentioned capacity restraints of the network. Indeed, the length of the training sequence grows as the binomial coefficient $\binom{n}{k}$ where k = 2, therefore for four and five targets (i.e., for n=4 or 5) the number of mixtures is increased to 6 and 10, respectively.

\subsection{Experiment 2: Sound Segregation with Alternating Multiple Mixtures of Synthesized Sounds }
\begin{figure}[t]
 \centering
     \includegraphics[width=0.75\textwidth]{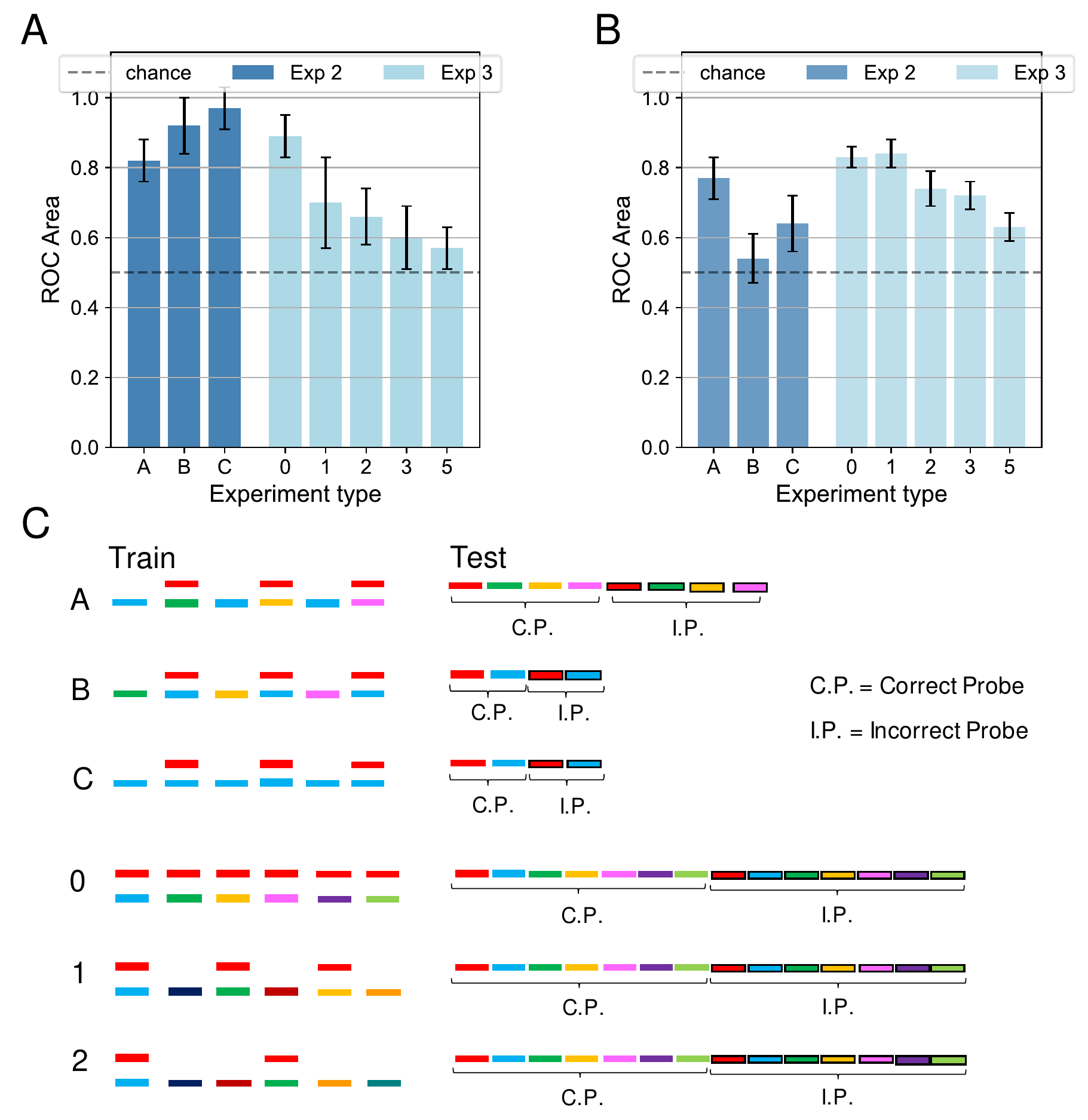}
\caption{{\bf Experiments 2 and 3 – results and comparison with human performance.}
(\textbf{A}) Results for Experiments 2 (dark blue) and 3 (light blue) on the dendritic network model. In Experiment 2 the performance is above chance for the three conditions. In Experiment 3 the accuracy decreases as the number of isolated sounds alternating with the mixtures increases. (\textbf{B}) Results for Experiments 2 (dark blue) and 3 (light blue) on the human experiment. In Experiment 2 the performance is above chance in the conditions A and C, while it is random for condition B. In Experiment 3 the accuracy decreases as the target presentation is more delayed. (\textbf{C}) Schematics for Experiments 2 and 3. The training is the same for both the dendritic network model and the human experiment. The schematics is omitted for delays 3 and 5. The testing refers to the dendritic network model, while the testing for the human experiment (same as in figure \ref{fig5}B) is omitted.}
\label{fig6}
\end{figure}

Next, we investigate the model's performance when the training sequence alternated mixtures of sounds with isolated sounds. Only the target and the masker sounds were later tested since recognizing the sounds presented individually during training would have been trivial. An analogous protocol was tested in psychophysical experiment (see experiment 3 in \cite{McDermott11}). Figures \ref{fig6}A and \ref{fig6}B show the network accuracy and human performance, respectively, for the protocols A,B,C in figure \ref{fig6}C. In the alternating task, the network was only partially able to reproduce the human results, displaying an interesting contrast to human behaviour. In condition A, in which the sounds mixed with the main target (in red) changed during training, the listeners were able to learn the targets with an accuracy of about 80\%, and so did our model. In contrast, our network behaved radically differently with respect to human performance under condition B, in which the training sequence consisted of the same mixture alternating with different sounds. As reported in figure \ref{fig5}B, the listeners were generally not able to identify the single sounds composing the mixture. Our model, instead, unexpectedly achieved a performance well above chance. The output dynamics could distinguish the distractors from the two targets with accuracy surprisingly above 90\%. 
The behavioural discrepancy under condition B could be explained by considering that in the training scheme the network is presented with three different sounds besides the mixture. With respect to Experiment 1 with a single mixture, in this protocol the network could learn the supplementary features of the isolated sounds and could exploit them during inference to respond differently to the distractors. From the spectrograms shown in figure \ref{fig2}, it is evident that some regions of overlap exist between the higher-intensity areas of different sounds. Therefore, the network presented during training with isolated sounds in addition to the single mixture, could detect some similarities between the training sounds and the tested distractors and respond with a more defined output dynamics than in Experiment 1. Finally, under condition C, both human subjects and our model performed above chance. While human performance was slightly above 60\%, the network achieved more than 90\% accuracy. This result should be interpreted considering that during inference also the isolated sound (blue) was tested together with the associated distractor, which was a trivial task for the nature of our network and thus boosted its overall performance. 

\subsection{Experiment 3: Effect of Temporal Delay in Target Presentation with Synthesized Sounds }
Temporal delay in the presentation of mixtures containing the target degraded performance similarly in the model and human subjects. We presented the network with a training sequence of six mixtures containing the same target mixed each time with a different distractor (figure \ref{fig6}C, protocols 0,1,2: c.f. experiment 4 in \cite{McDermott11}). The mixtures alternated with an increasing number of isolated sounds, hence increasing the interval between successive presentations of the target. The human ability to extract single sounds from mixtures was previously shown to worsen as the interval between target presentations increased, as replicated in figure \ref{fig6}B. The network presented a similar decreasing trend, as reported in figure \ref{fig6}A. An interesting difference, however, is that the performance of our model drastically dropped even with one isolated sound every other mixture while the human performance was affected when at least two isolated sounds separated the target-containing mixtures. The discrepant behaviour indicates that the insertion of isolated sounds between the target-containing mixtures more strongly interferes the learning of the target sound in the model compared to human subjects. This stronger performance degradation may partly be due to the capacity constraint of our simple neural model, which uses a larger amount of memory resource as the number of isolated sounds increases. In contrast, such a constraint may be less tight in the human auditory systems. 

Also for Experiments 2 and 3, we tested a modification of the inference protocol, by presenting the network only with the target sound and one incorrect probe. Under this configuration, the model performance of Experiment 2 improves compared to the original protocol, while no substantial changes are noted for Experiment 3 (Supplementary figure 3).

\subsection{Experiment 4: Sound Segregation with Single and Multiple Mixtures of Real-World Sounds }
\begin{figure}[t]
 \centering
     \includegraphics[width=1\textwidth]{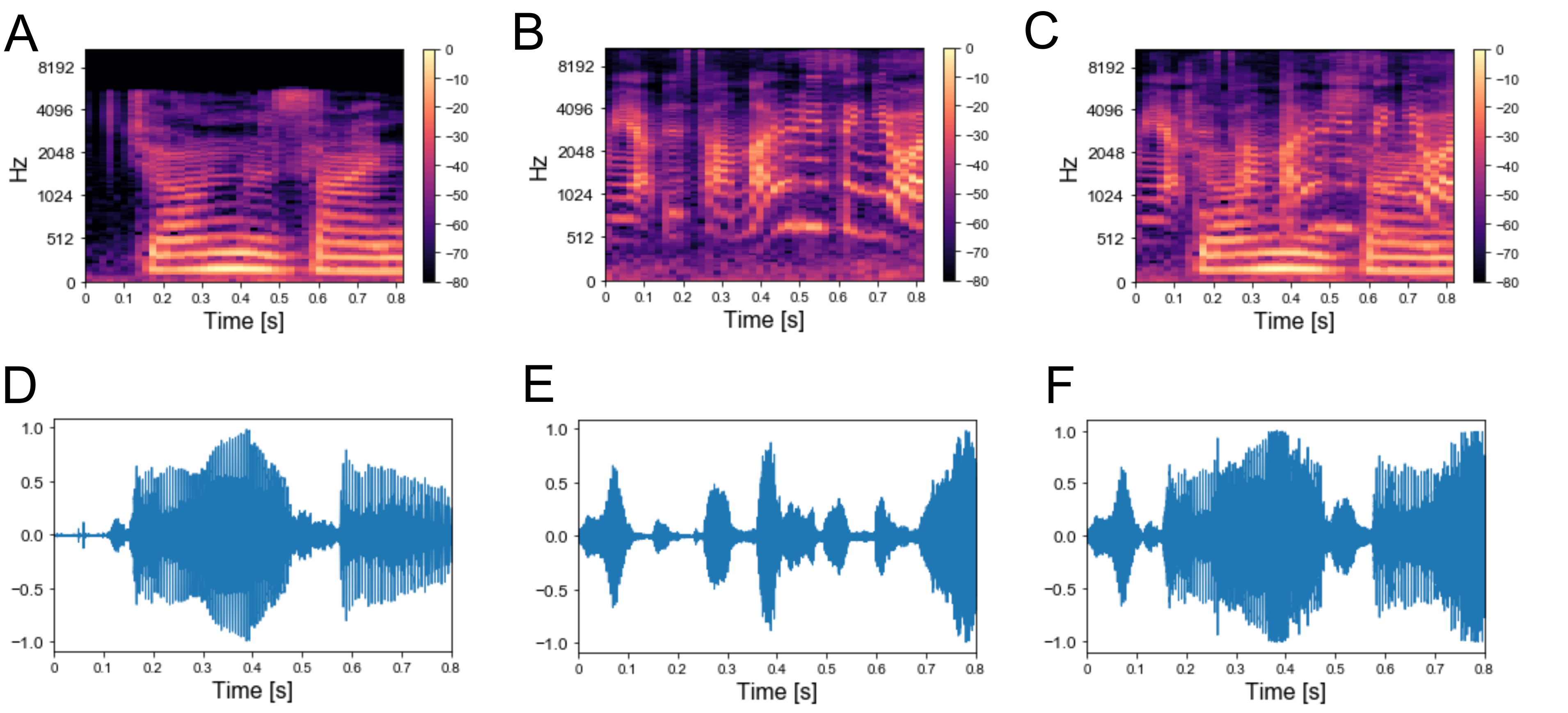}
\caption{{\bf Real-world sounds – Targets and mixture.}
(\textbf{A}) Spectrogram of a spoken sentence 800 ms-long. (\textbf{B}) Spectrogram of 800 ms-long recording of chimes sounds. (\textbf{C}) Spectrogram of the mixture of the sounds in A and B. (\textbf{D}) Sound wave associated with the spectrogram in A. (\textbf{E}) Sound wave associated with the spectrogram in B. (\textbf{F}) Sound wave associated with the spectrogram in C.}
\label{fig7}
\end{figure}

\begin{figure}[t]
 \centering
     \includegraphics[width=1\textwidth]{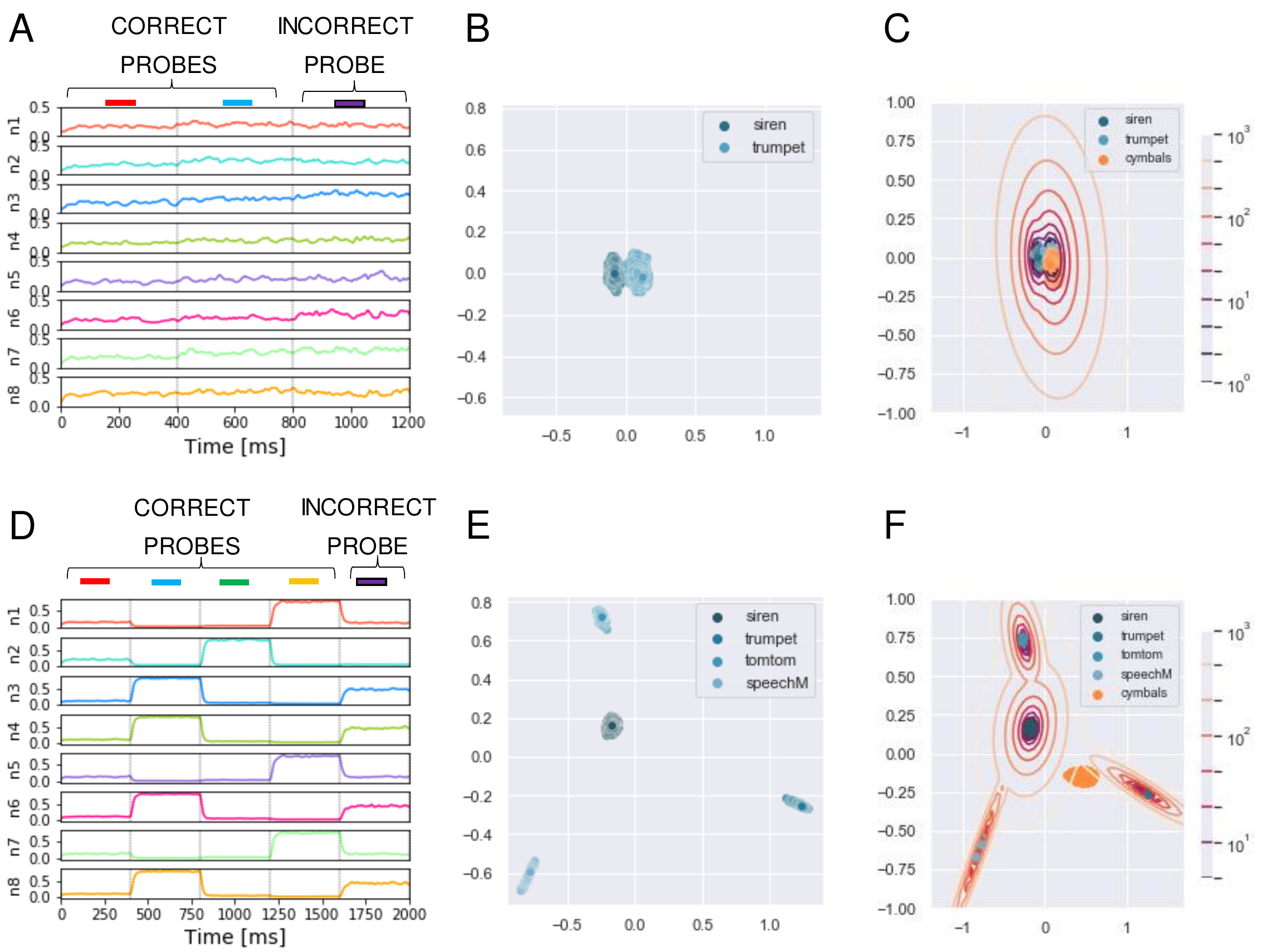}
\caption{{\bf Experiment 4 - output dynamics and clustering.}
(\textbf{A,B,C}) refer to the results of Experiment 4 on real-world sounds with a single mixture presented during training. (\textbf{D,E,F}) refer to the results of Experiment 4 on real-world sounds with three mixtures presented during training. (\textbf{A}) Voltage dynamics of the 8 output neurons during inference, when the target, the distractor and one unseen sound are tested. As expected, the neuron population is not able to respond with different dynamics to the three sounds, and the voltage of all the output neurons fluctuates randomly throughout the whole testing sequence. (\textbf{B}) The PCA projection of the datapoints belonging to the target and distractor (in blue) shows that the clusters are collapsed into a single cluster. (\textbf{C}) When GMM is applied, all the datapoints representing both the learnt sounds (in blue) and the unseen sound (in orange) fall within the same regions, making it impossible to distinguish the different sounds based on the population dynamics. (\textbf{D}) Voltage dynamics of the 8 output neurons during inference, when the target, the three distractors, and one unseen sound are tested. As expected, the neuron population has learnt the feature of the different sounds and responds with different dynamics to the five sounds. Each output neuron has an enhanced response to one or few sounds. (\textbf{E}) The PCA projection of the datapoints belonging to the four correct probes (in blue) shows that the clusters are more compact and more spatially distant one from the other with respect to the result obtained with the synthetized sounds. (\textbf{F}) When GMM is applied, the model shows that the network clearly distinguished the learnt sounds (in blue) from the unseen sound (in orange). These results show that the grouping cues improve the model accuracy with respect to the synthesized dataset.}
\label{fig8}
\end{figure}

We applied the same protocol of Experiments 1 to the dataset of natural sounds. Although such experiments were previously not attempted on human subjects, it is intriguing to investigate whether the model can segregate target natural sounds by the same strategy. The spectrograms of two isolated sounds and of their mixtures are shown in figure \ref{fig7}A-C, together with the respective sound waves (figure \ref{fig7}D-F). The qualitative performance was very similar to that obtained with the synthesized sounds. Specifically, the output dynamics learned from the repetition of a single mixture was randomly fluctuating for both seen and randomly chosen unseen sounds (figure \ref{fig8}A), whereas the network responses to targets and unseen sounds were clearly distinct if multiple mixtures were presented during training (figure \ref{fig8}D). The output dynamics were not quantitatively evaluated because it was not possible to rigorously generate incorrect probes associated with the learnt targets and distractors. Therefore, we qualitatively assessed the performance of the model by observing the clustering of network responses to the learnt targets versus unseen natural sounds (figure \ref{fig8}B, C, E and F). We observed that, in the case of multiple mixtures, the clusters related to natural sounds (figure \ref{fig8}E and F) were more compact than those of synthetic sounds (figure \ref{fig4}E and F). Furthermore, these clusters were more widely spaced on the PCA projection plane: the intraclass correlation in the response to the same target was greater while the interclass similarity in the response to different targets or distractors was lower. These results indicate that grouping cues, such as harmonic structure and temporal onset, improve the performance of the model.

\subsection{Experiment 5: Image Segregation with Single and Multiple Mixtures of Real-World Images }
\begin{figure}[t]
 \centering
     \includegraphics[width=1\textwidth]{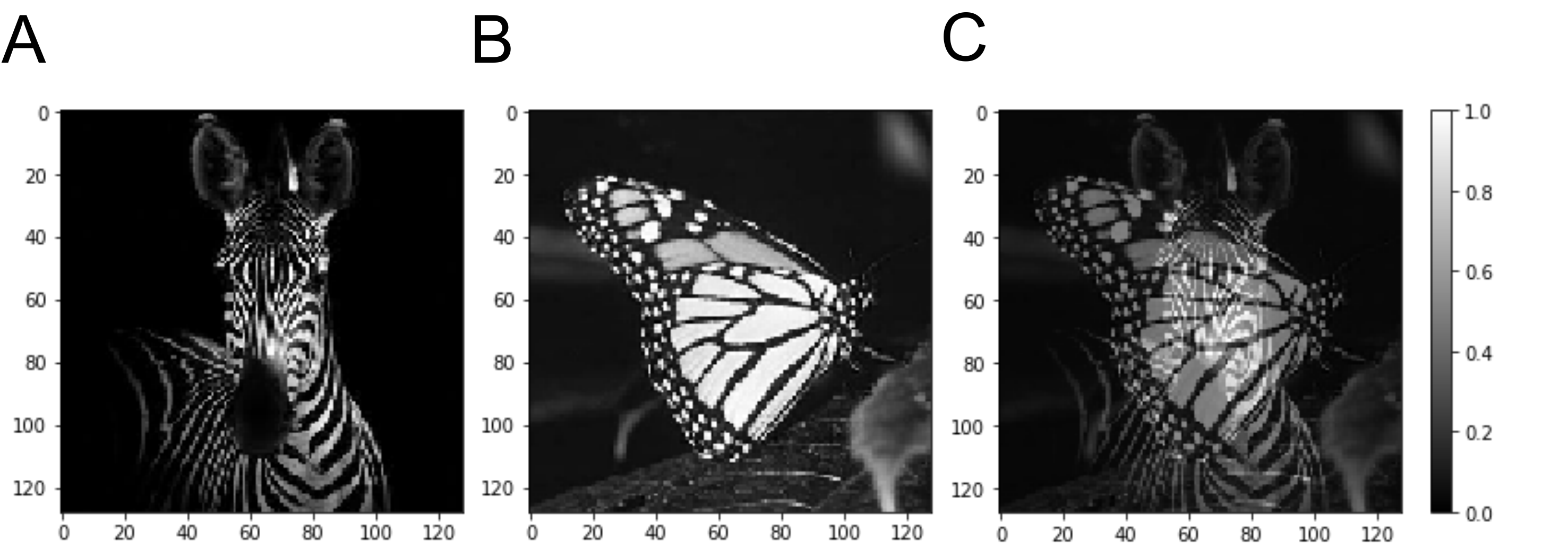}
\caption{{\bf Real-world images – Targets and mixture.}
(\textbf{A}) Squared 128x128 target image of a zebra. (\textbf{B}) Squared 128x128 distractor image of a butterfly. (\textbf{C}) Mixture of the target and distractor images shown in A and B.}
\label{fig9}
\end{figure}

\begin{figure}[t]
 \centering
     \includegraphics[width=1\textwidth]{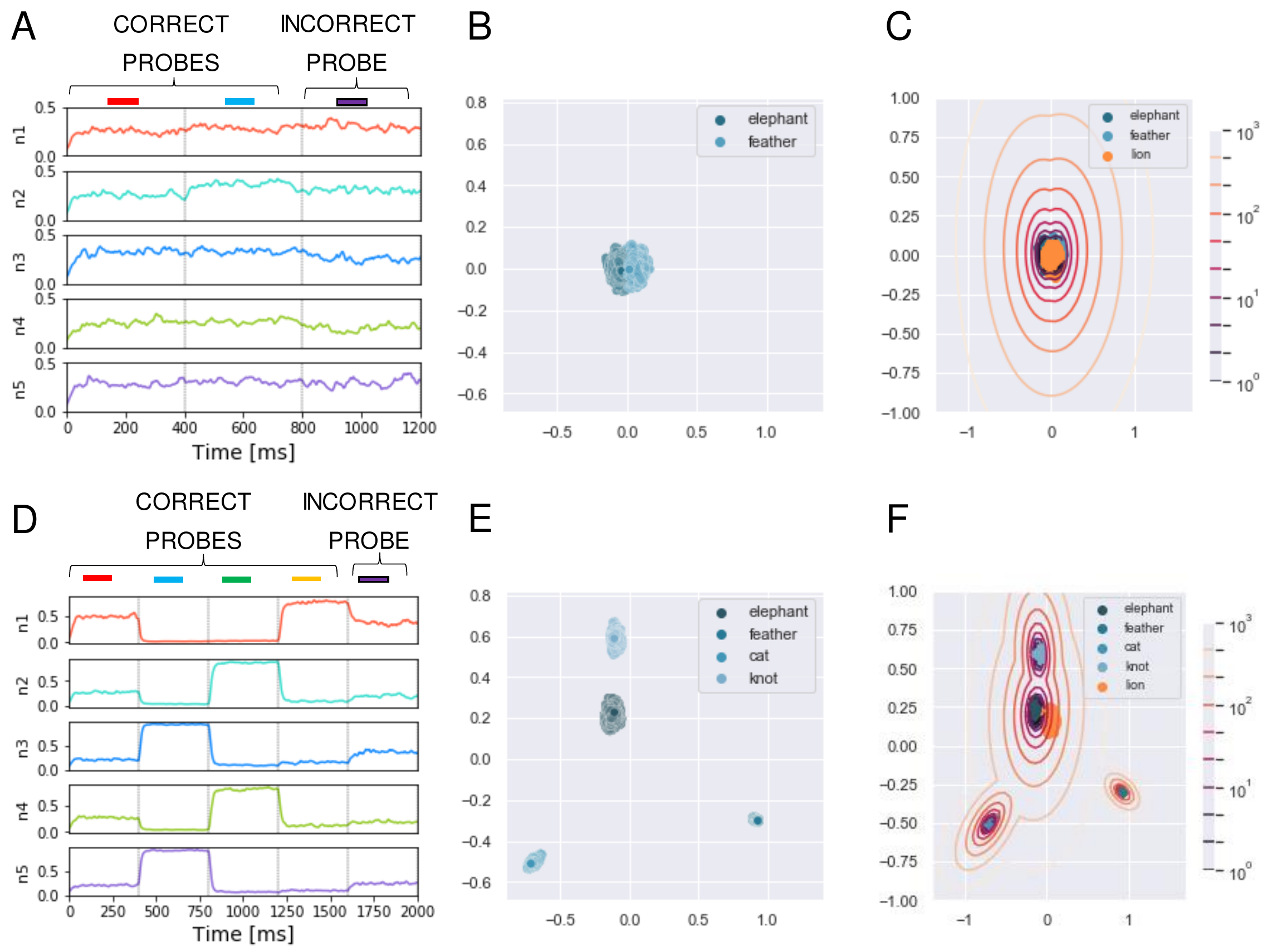}
\caption{{\bf Experiment 5 – output dynamics and clustering.}
(\textbf{A,B,C}) refer to the results of Experiment 5 on real-world images with a single mixture presented during training. (\textbf{D,E,F}) refer to the results of Experiment 5 on real-world images with three mixtures presented during training. (\textbf{A}) Voltage dynamics of the 5 output neurons during inference, when the two training images and one unseen image are tested. As expected, the neuron population is not able to respond with different dynamics to the three images, and the voltage of all the output neurons fluctuates randomly throughout the whole testing sequence. (\textbf{B}) The PCA projection of the datapoints belonging to the two seen images (in blue) shows that the clusters are collapsed into a single cluster. (\textbf{C}) When GMM is applied, all the datapoints representing both the targets (in blue) and the unseen image (in orange) fall within the same regions, making it impossible to distinguish the different images based on the population dynamics. (\textbf{D}) Voltage dynamics of the 5 output neurons during inference, when the four targets and one unseen image are tested. As expected, the neuron population has learnt the features of the different images and responds with different dynamics to the five images. Each output neuron has an enhanced response to one or few inputs. (\textbf{E}) The PCA projection of the datapoints belonging to the four learnt images (in blue) shows that the clusters are compact and spatially distant one from the other. (\textbf{F}) When GMM is applied, the model shows that the network clearly distinguished the target and distractors (in blue) from the unseen image (in orange). These results suggest that humans would be able to distinguish single visual targets if previously seen in different mixtures.}
\label{fig10}
\end{figure}

Finally, we examined whether the source segregation through repetition scheme can also extend to vision-related tasks, as previously suggested \cite{McDermott11}. To this end, we employed the same method as developed for sound sources and performed the recovery of visual sources with the protocol of Experiment 1. The mixtures were obtained by overlapping black-and-white images sampled from our visual dataset (Material and Methods), as shown in figure \ref{fig9}. We remark that the model is presented with the visual stimuli following the same computational steps as for sounds. Indeed, as previously described, the acoustic stimuli are first pre-processed into spectrograms and then encoded by the input layer.  Similarly to Experiment 4, the performance of the model was assessed only qualitatively in the visual tasks. As in the acoustic tasks, the clustering of network responses showed that the model was able to retrieve the single images only when more than one mixture was presented during training. The network responses are shown in figure \ref{fig10}.

\section{Discussion}
The recovery of individual sound sources from mixtures of multiple sounds is a central challenge of hearing. Based on experiments on human listeners, sound segregation has been postulated to arise from prior knowledge of sound characteristics or detection of repeating spectro-temporal structure. The results of \cite{McDermott11} show that a sound source can be recovered from a sequence of mixtures if it occurs more than once and is mixed with more than one masker sound. This supports the hypothesis that the auditory system detects repeating spectro-temporal structure embedded in mixtures, and interprets this structure as a sound source.
We investigated whether a biologically inspired computational model of the auditory system can account for the characteristic performance of human subjects. To this end, we implemented a one-layer neural network with dendritic neurons followed by a readout layer based on GMM to classify probe sounds as seen or unseen in the training mixtures. 
The results in \cite{McDermott11} show that source repetition can be detected by integrating information over time and that the auditory system can perform sound segregation when it is able to recover the target sound's latent structure. Motivated by these findings, we trained our dendritic model with a learning rule that was previously demonstrated to detect and analyze the temporal structure of a stream of signals. In particular, we relied on the learning rule described by \cite{asabuki20}, which is based on the minimization of regularized information loss. Specifically, such a principle enables the self-supervised learning of recurring temporal features in information streams using a family of competitive networks of somatodendritic neurons. 

We presented the network with temporally overlapping sounds following the same task protocols as described in \cite{McDermott11}. First, we carried out the segregation task with the same dataset of synthesized sounds presented to human listeners in \cite{McDermott11}.
We found that the model was able to segregate sounds only when one of the masker sounds varied, not when both sounds of the mixture were repeated. Our findings bear a closer resemblance to the experimental findings of human listeners over a variety of task settings. 
Earlier works have proposed biologically inspired networks to perform BSS \cite{Isomura19,bahroun21,pehlevan17}. However, to our knowledge, this is the first attempt to reproduce the experimental results of recovering sound sources through embedded repetition. Additionally, we demonstrated that our network can be a powerful tool for predicting the dynamics of brain segregation capabilities under settings difficult to test on humans. In particular, the recovery of natural sounds is expected to be a trivial task for humans given their familiarity with the sounds, whereas our model is built from scratch and has no prior knowledge about natural sounds. We find that the hallmarks of natural sounds make the task easier for the network when the target is mixed with different sounds, but, as for the synthetic dataset, the sounds cannot be detected if presented always in the same mixture. Furthermore, we extended the study to investigate BSS of visual stimuli and observed a similar qualitative performance as in the auditory settings. This is not surprising from a computational perspective as the computational steps of the visual experiment are the same as for the acoustic experiment: there, the sounds are first preprocessed into images, the spectrograms, and then presented to the network in a visual form. From the biological point of view, the neural computational primitives used in the visual and the auditory cortex may be similar, as evidenced by anatomical similarity and by developmental experiments where auditory cortex neurons acquire V1-like receptive fields when visual inputs are redirected there \cite{sharma00,bahroun21}. 
Motivated by these reasons, we suggest extending the experiments of source repetition to vision to verify experimentally whether our computational results provide a correct prediction of the source separation dynamics of the visual system.

Although the dynamics of our model under many aspects matches the theory of repetition-based BSS, the proposed scheme presents a few limitations. The major limitation concerns the discrepancy of the results in experiment 2B. In such a setting, the model performance is well above chance, although the target sound always occurs in the same mixture. We speculate that, in this task settings, the output neurons learn the temporal structure of the distractor sounds presented outside the mixture and that they recognize some similarities in the latent structure of the probes. This pushes the neurons to respond differently to the correct and incorrect probes, thereby allowing the output classifier to distinguish the sounds. In contrast, we speculate that human auditory perception relies also on the outcome of the later integration of features detected at early processing stages. This will prevent the misperception of sounds based on unimportant latent features. A second limitation of the selected encoding method consists in the difficulty to model the experiments relying on the asynchronous overlapping of signals and on reversed probe sounds presented by \cite{McDermott11}. Indeed, in our approach, each input neuron responds to one specific time frame, and the output neurons are trained uniquely on this configuration. Hence, temporal shifts or inverting operations are not possible.

Furthermore, as previously mentioned, the training scheme in \cite{asabuki20} has proven to be able to learn temporal structures in a variety of tasks. In particular, the model was shown to perform chunking as well as to achieve BSS from mixtures of mutually correlated signals. 
We underline that our computational model and experiments differ in fundamental ways from the BSS task described by \cite{asabuki20}. First, the two experiments diverge in their primary scope. The BSS task aims at using the average firing rate of the single neurons responding to sound mixtures to decode separately the original sounds. In our work, instead, sound mixtures are included only in the training sequence and, during inference, only individual sounds are presented to the network. Our goal is to verify from the population activity whether the neurons have effectively learnt the sounds and can distinguish them from unseen distractors. Furthermore, in \cite{asabuki20} the stimulus was encoded into spike patterns using one Poisson process proportional to the amplitude of the sound waveform at each time step, disregarding the signal intensity at different frequencies. This method was not suitable for the source segregation through repetition task, where the sound mixtures retain important information on the frequency features of the original sounds at each time frame.

In summary, we have shown that a network of dendritic neurons trained in an unsupervised fashion is able to learn the temporal structure of overlapping sounds and, once the training is completed, can perform blind source separation if the individual sounds have been presented in different mixtures. These results account for the experimental performance of human listeners tested in the same task setting. Our study has demonstrated that a biologically inspired simple model of the auditory system can capture the intrinsic neural mechanisms underlying the brain's capability of recovering individual sound sources based on repetition protocols.

\section{Materials and Methods}
\subsection{Datasets}
A dataset of synthesized sounds were created in the form of spectrogram, which shows how signal strength evolves over time at various frequencies, according to the method described previously \cite{McDermott11}. In short, the novel spectrograms were built as Gaussian distributions based on correlation functions analogous to those of real-world sounds. White noise was later applied to the resulting spectrograms. Five Gaussian distributions were employed to generate each of ten different sounds in figure \ref{fig5}A. The corresponding spectrograms featured 41 frequency filters equally spaced on an ERBN (Equivalent Rectangular Bandwidth, with subscript N denoting normal hearing) scale \cite{Glasberg90} spanning 20-4000 Hz, and 31 time frames equally dividing the 300ms sound length. For our simulations, we used the same MATLAB toolbox and parameters used in the previous study \cite{McDermott11}. For further details on the generative model for sounds, please refer to the SI Materials and Methods therein.

In addition to the dataset of synthesized sounds, we built a database composed of 72 recordings of isolated natural sounds. The database contained 8 recordings of human speech from the EUSTACE (the Edinburgh University Speech Timing Archive and Corpus of English) speech corpus \cite{eustace03}, 23 recordings of animal vocalizations from the Animal Sound Archive \cite{Frommolt06}, 29 recordings of music instruments by Philharmonia Orchestra \cite{philharmonia}, and 12 sounds produced by inanimate objects from the BBC Sound Effect corpus \cite{BBCsoundeffects91}. The sounds were cut into 800ms extracts. Then the library librosa \cite{McFee15} was employed to extract spectrograms with 128 frequency filters spaced following the Mel scale \cite{Stevens37} and 10ms time frames with 50\% overlap. 

For image source separation, we built a database consisting of 32 black-and-white pictures of various types, both single objects and landscapes. The images were later squared, and their size was reduced to 128x128 pixels.

\subsection{Neuron model}
In this study we used the same two-compartment neuron model as that developed previously \cite{asabuki20}. The mathematical details are found therein. Here, we only briefly outline the mathematical framework of the neuron model. Our two-compartment model learns temporal features of synaptic input given to the dendritic compartment by minimizing a regularized information loss arising in signal transmission from the dendrite to the soma. In other words, the two-compartment neuron extracts the characteristic features of temporal input by compressing the high dimensional data carried by a temporal sequence of presynaptic inputs to the dendrite onto a low dimensional manifold of neural dynamics. The model performs this temporal feature analysis by modifying the weights of dendritic synapses to minimize the time-averaged mismatch between the somatic and dendritic activities over a certain recent interval. In a stationary state, the somatic membrane potential of the two-compartment model could be described as an attenuated version of the dendritic membrane potential with an attenuation factor \cite{Urbanczik14}. Though we deal with time-dependent stimuli in our model, we compare the attenuated dendritic membrane potential with the somatic membrane potential at each time point. This comparison, however, is not drawn directly on the level of the membrane potentials but on the level of the two non-stationary Poissonian spike distributions with time-varying rates, which would be generated if both soma and dendrite were able to emit spikes independently. In addition, the dynamic range of somatic responses needs to be appropriately rescaled (or regularized) for meaningful comparison. An efficient learning algorithm for this comparison can be derived by minimizing the Kullback–Leibler (KL) divergence between the probability distributions of somatic and dendritic activities. Note that the resultant learning rule enables unsupervised learning because the somatic response is fed back to the dendrite to train dendritic synapses. Thus, our model proposes the view that backpropagating action potentials from the soma may provide a supervising signal for training dendritic synapses \cite{Larkum13,Larkum99}.

\subsection{Network architecture}
The network architecture, shown in figure \ref{fig1}, consisted of two layers of neurons, either fully connected or with only 30\% of the total connections. The input layer contained as many Poisson neurons as the number of pixels present in the input spectrogram (acoustic stimulus) or input image (visual stimulus). The postsynaptic neurons were modelled according to the two-compartment neuron model proposed previously \cite{asabuki20}. Their number was varied from a pair to few tenths, depending on the complexity of the task. Unless specified otherwise, 8 and 5 output neurons were set for acoustic and visual task respectively. 

In the first layer, the input was encoded into spikes through a rate coding-based method \cite{Almomani19}. The strength of the signal at each pixel drove the firing rate of the associated input neuron, i.e. the spike trains were drawn from Poisson processes with probability proportional to the intensity of the pixel. We imposed that, for each input stimulus, the spike pattern was generated through a sequence of 400 Poisson processes.

We designed the output layer and the learning process similarly to the previous network used for the blind signal separation (BSS) within mixtures of multiple mutually correlated signals as well as for other temporal feature analyses \cite{asabuki20}. As mentioned previously, the learning rule was modelled as a self-supervising process, which is at a conceptual level similar to Hebbian learning with backpropagating action potentials. The soma generated a supervising signal to learn and detect the recurring spatiotemporal patterns encoded in the dendritic activity. Within the output layer, single neurons learned to respond differently to each input pattern. Competition among neurons was introduced to ensure that different neurons responded to different inputs. With respect to the network used for BSS containing only two output neurons, we rescaled the strength of the mutual inhibition among dendritic neurons by a factor proportional to the inverse of the square root of the number of output neurons. This correction prevented each neuron from being too strongly inhibited when the size of the output layer increased (i.e. exceeds three or four). Furthermore, we adopted the same inhibitory spike timing-dependent plasticity (iSTDP) as employed in the previous model. This rule modified inhibitory connections between two dendritic neurons when they coincidently responded to a certain input.  The iSTDP allowed the formation of chunk-specific cell assemblies when the number of output neurons was greater than the number of input patterns. 

For all parameters but noise intensity during learning, we used the same values as used in the original network model \cite{asabuki20}. For bigger values of g, the neural responses were subject to more fluctuations and neurons tended to group in only one cell assembly. From the analysis of the learning curves shown in figure \ref{fig3}, we decided to train the network from randomly initialized weights and to expose it, during training, to the mixture sequence 3000 times for the synthesized sounds and 1500 times for the real-world sounds. The learning rate was kept constant throughout the whole process. During testing, the sequence of target sounds and respective distractors was presented 50 times, and the resulting neural dynamics was averaged over 20 trials. The performance results shown in the Results section were computed as average over 10 repetitions of the same simulation set-up. In each repetition different target sounds and distractors were randomly sampled from the dataset in order to ensure performance independence of specific sounds.

\subsection{Experimental settings and performance measure}
The synapses were kept fixed during inference in our network, implying that the responses to probes tested later were not affected by the presentation of other previously tested probes. This allowed us to test the trained network on a sequence of probes, rather than only on one probe as in the studies of the human brain where plasticity cannot be frozen during inference \cite{McDermott11}. In figure \ref{fig5}A and figure \ref{fig6}C, the first half of the sequence contained the target and the distractors, the second half the respective incorrect probes, which were also built by using the same method as in human experiment \cite{McDermott11}. Each incorrect probe was a sound randomly selected from the same Gaussian distribution generating the associated target. After the sampling, a randomly selected time slice equal to 1/8 of the sound duration was set to be equal to the target. 

The possibility of presenting more than one probe allowed us to test the performance of the network for all the sounds present in the mixtures. To ensure a stable neural response against the variability of the encoding, we repeated the sequence 50 times. The response of the network consisted of the ensemble activity of the output neurons. As previously explained, 400 time-steps were devoted to the presentation to each stimulus. The response to each probe, therefore, consisted of 400 data points describing the dynamical activity of each output neuron, each point being a collection of N values, where N is the number of output neurons. An example of one testing epoch output is shown in figure \ref{fig4}A and figure \ref{fig4}C. We neglected the first 50 data points, since, during the initial transient time, the membrane potential was still decaying or rising after the previous input presentation. For visualization purpose, we applied the principal component analysis (PCA) to reduce the dimensionality of the data from N to 2. In our settings, the two principal components explain approximately 40\% of the variance of the neural response. The PCA transformation was based uniquely on the data points obtained with the presentation of the target and the distractors, as shown in figure \ref{fig4}B and Figure figure \ref{fig4}E. The same transformation was later exploited to project the points related to the incorrect probes. Only the target and distractors patterns were presented during the learning process, and the responses to unseen patterns were afterwards projected on the space defined by the training. 

The two-dimensional projection of the target-related data points were clustered in an unsupervised manner through GMM. We set the number of Gaussians equal to the number of targets such that the covariance matrices had a full rank. With the defined GMM model at hand, we proceeded with fitting all the PCA data points, related to both correct and incorrect probes. The model tells which cluster each data point belonged to and what was the likelihood (L) that the cluster had generated this data point. figure \ref{fig4}C and figure \ref{fig4}F show the datapoints projected on the PCA plane together with the GMM clustering and likelihood curves. 

We used the likelihood as a measure of performance. The four intervals of the likelihood range corresponding to the four responses “sure no”, “no”, “yes”, and “sure yes” were (i) $L > 0$ (sure yes), (ii) $-5 < L < 0$ (yes), (iii) $-15 < L < -5$ (no), and (iv) $L < -15$ (sure no). In building the receiver operating characteristic (ROC) curve, the datapoints falling in the interval (i) were assigned the probability value 1.0, those in (ii) 0.66, those in (iii) 0.33, and those in (iv) 0.0.

The described evaluation metrics was applied only to the experiments carried on the dataset composed of synthesized sounds. For the experiments based on natural sounds and images, the results of clustering were shown only qualitatively for the target-related datapoints. Indeed, due to the real-world nature of signals, it was not possible to simply use Gaussian functions to build physically consistent incorrect probes. On the real-world sound dataset, we performed all the same protocol of Experiment 1 (Experiment 4).
On the image dataset we performed an experiment with a protocol analogous to Experiment 1. Here, the mixtures were obtained by overlapping two images, both with transparency 0.5, similarly to the spectrogram overlapping described for the acoustic task. The input images were normalized to the range [0,1] and the intensity of each pixel was encoded through the firing rate of one input neuron. We followed the same procedure and network setting described for the audio stimuli segregation to assess the ability of the network to separate visual stimuli presented in mixtures.

\section*{Conflict of Interest Statement}
%All financial, commercial or other relationships that might be perceived by the academic community as representing a potential conflict of interest must be disclosed. If no such relationship exists, authors will be asked to confirm the following statement: 
The authors declare that the research was conducted in the absence of any commercial or financial relationships that could be construed as a potential conflict of interest.

\section*{Author Contributions}
T.F., G.D. and T.A. conceived the idea. G.D. designed and performed the simulations, with input from T.A.. All authors analyzed the results. G.D. and T.F. wrote the manuscript. T.A. and G.D. wrote the Supplementary material.

\section*{Funding}
This work was partly supported by JSPS KAKENHI no. 19H04994 to T.F.

\section*{Acknowledgments}
We are grateful to all the colleagues in the Neural Coding and Brain Computing Unit for fruitful interaction.

\section{Supplemental Data}
% \href{http://home.frontiersin.org/about/author-guidelines#SupplementaryMaterial}{Supplementary Material} should be uploaded separately on submission, if there are Supplementary Figures, please include the caption in the same file as the figure. LaTeX Supplementary Material templates can be found in the Frontiers LaTeX folder.
\paragraph*{S1 Appendix.}
\label{S1_Appendix}
{\bf Appendix.} Equations describing the two-compartment neuron model and synaptic learning rule.

\paragraph*{S1 Figure}
\label{S1_Fig}
{\bf Experiment 1 with varying N and p.} Results for Experiment 1 for different numbers of output neurons (N) and different connectivity probabilities (p).

\paragraph*{S2 Figure}
\label{S2_Fig}
{\bf Experiment 1 with target-based test.} Results for Experiment 1 using only the correct and incorrect probes related to the target sound during inference.

\paragraph*{S3 Figure}
\label{S3_Fig}
{\bf Experiments 2 and 3 with target-based test.} Results for Experiments 2 and 3 using only the correct and incorrect probes related to the target sound during inference.

\section*{Data Availability Statement}
The dataset of synthesized sounds analyzed in this study can be found in the section \textit{Gaussian Sound Synthesis Toolbox} [\url{https://mcdermottlab.mit.edu/downloads.html}].\\
The datasets of natural sounds created for this study can be found at the repository \url{https://github.com/GiorgiaD/dendritic-neuron-BSS}.

%Bibliography
\bibliographystyle{unsrt}  
\bibliography{references}

\end{document}